\documentclass[proof]{pasj00}
\draft

\begin{document}
\SetRunningHead{Author(s) in page-head}{Running Head}

\title{Dependence of the Magnetic Energy of Solar Active Regions
on the Twist Intensity of the Initial Flux Tubes}

%

%
 \author{%
   Shin~\textsc{Toriumi}\altaffilmark{1}
   Takehiro~\textsc{Miyagoshi}\altaffilmark{2}
   Takaaki~\textsc{Yokoyama}\altaffilmark{1}
   Hiroaki~\textsc{Isobe}\altaffilmark{3}\\
   and
   Kazunari~\textsc{Shibata}\altaffilmark{4}}
 \altaffiltext{1}{Department of Earth and Planetary Science,
   University of Tokyo,\\ Hongo, Bunkyo-ku, Tokyo 113-0033}
 \email{toriumi@eps.s.u-tokyo.ac.jp}
 \altaffiltext{2}{Japan Agency for Marine-Earth Science and Technology,\\
   Showa-machi, Kanazawa-ku, Yokohama 236-0001}
 \altaffiltext{3}{Unit of Synergetic Studies for Space,
Kyoto University, Yamashina-ku, Kyoto 607-8471}
 \altaffiltext{4}{Kwasan and Hida Observatories, Kyoto University,
Yamashina-ku, Kyoto 607-8471}


\maketitle

\begin{abstract}
We present a series of numerical experiments
that model the evolution of magnetic flux tubes
with a different amount of initial twist.
As a result of calculations,
tightly twisted tubes reveal
a rapid two-step emergence to the atmosphere
with a slight slowdown at the surface,
while weakly twisted tubes show
a slow two-step emergence
waiting longer the secondary instability
to be triggered.
This picture of the two-step emergence
is highly consistent with recent observations.
These tubes show multiple magnetic domes
above the surface,
indicating that the secondary emergence
is caused by interchange mode
of magnetic buoyancy instability.
As for the weakest twist case,
the tube exhibits an elongated photospheric structure
and never rises into the corona.
The formation of the photospheric structure
is due to inward magnetic tension force
of the azimuthal field component
of the rising flux tube
(i.e., tube's twist).
When the twist is weak,
azimuthal field cannot hold the tube's coherency,
and the tube extends laterally
at the subadiabatic surface.
In addition,
we newly find that the total magnetic energy
measured above the surface depends on the initial twist.
Strong twist tubes follow the initial relation
between the twist and the magnetic energy,
while weak twist tubes deviates from this relation,
because these tubes store their magnetic energy
in the photospheric structures.
\end{abstract}

\section{Introduction\label{sec:intro}}

Flux emergence is
one of the key mechanisms
in various solar activities.
It is widely accepted
that the emerging flux has
a form of a twisted flux tube
so as not to be collapsed
by the convective motions
during its ascent
in the solar interior.
Emerging flux transports
magnetic energy and helicity
from the convection zone
to the atmosphere,
which yields active regions
including sunspots.
Magnetic helicity in the corona
stores free energy
that can be released
in the forms of flares
and coronal mass ejections (CMEs)
(e.g. \cite{hey77}).

Many numerical experiments
have revealed the dynamics
of the flux emergence.
\citet{sch79} carried out
two-dimensional magnetohydrodynamic (MHD)
simulations to study
the cross-sectional evolution
of the emerging flux tube
(see also \cite{mor96,emo98}).
\citet{shi89} calculated
the two-dimensional evolution of
the undular mode of magnetic buoyancy instability
(Parker instability: \cite{par66})
to reproduce the formation of
an $\Omega$-shaped coronal loops.
\citet{tor10a} and \citet{tor10b}
gave numerical studies
of emerging fluxes
from much deeper convection zone
($\sim -20,000$\ km) to the corona.
The three-dimensionality also exerts 
an influence on emerging process
of magnetic flux evolution.
\citet{mat93} produced
the first three-dimensional work
of the Parker instability
using a magnetic flux sheet
and a flux tube.
\citet{fan01} compared
her numerical results
of the twisted tube's emergence 
with observations of an active region.

In this paper,
we perform three-dimensional simulations
of the twisted emerging flux tube
from the uppermost convection zone
to the corona.
Our aim is to study
the effect of the initial twist
on the emergence process.
A series of parametric studies
on the flux tube's twist
was done by \citet{mur06}.
Our work is dedicated to further detailed analyses,
especially focusing
on the effect of the initial twist
on the resulting tube's structure
(photospheric lateral expansion
and multiple magnetic domes)
and on the consequent coronal magnetic energy.

For numerical experiments,
we used the same conditions
as those by \citet{mur06};
we calculated ten cases
of different twist parameters
that cover their three runs.
As a result of experiments,
we found that the evolution
depends on the initial twist.
When the twist is strong enough,
the evolution to the corona
reveals two-step way,
showing a deceleration
and a lateral expansion
near the solar surface,
although the case with weaker twist
spends more time waiting
for the secondary emergence to occur
\citep{mag01,arc04,mur06}.
This picture of the two-step emergence
is highly consistent
with recent observations by \citet{ots10},
especially its horizontally expanding speed
and the rising speed.
In addition to the confirmation
of the results by \citet{mur06},
it is also found
that multiple magnetic domes are built
and plasma accumulates in between the domes
when the secondary emergence starts.
At this moment,
the direction of the field lines
is almost perpendicular
to the alignment of the domes,
indicating that the second-step emergence
is due to the interchange-mode instability.
If the initial twist is too weak,
the tube extends widely near the surface
and further evolution never takes place,
because the magnetic tension force
of the azimuthal component
cannot hold the tube's coherency.

Also, we newly found that
the total magnetic energy
measured above the surface
relies on the initial twist.
In the strong twist regime,
the resulting magnetic energy follows
the initial relation
between the twist and the magnetic energy.
In the weak twist regime, however,
the magnetic energy deviates from
the initial rule,
because the tube with weak twist
stores magnetic energy
around the photosphere.

The rest of the paper
is organized as follows.
In Section \ref{sec:setup},
we describe the numerical model.
The simulation results
are shown in Section \ref{sec:results}.
Summary and discussion
are given in Section
\ref{sec:summary} and \ref{sec:discussion},
respectively.

\section{Numerical Setup\label{sec:setup}}

In numerical simulations,
we solve nonlinear, time-dependent, compressible
three-dimensional MHD equations.
We take a rectangular computation box
with three-dimensional Cartesian coordinates
($x,\, y,\, z$),
where the $z$-coordinate increases upward.
The medium is assumed to be
an inviscid perfect gas
with a specific heat ratio $\gamma =5/3$.
The basic equations
in vector form are as follows:
\begin{eqnarray}
  \frac{\partial\rho}{\partial t}
  + \mbox{\boldmath $\nabla$}
    \cdot(\rho\mbox{\boldmath $V$})=0,
\end{eqnarray}
\begin{eqnarray}
  \frac{\partial}{\partial t}
  (\rho\mbox{\boldmath $V$})
  + \mbox{\boldmath $\nabla$}\cdot
  \left(
    \rho\mbox{\boldmath $V$}\mbox{\boldmath $V$}
    +p\mbox{\boldmath $I$}
    -\frac{\mbox{\boldmath $BB$}}{4\pi}
    +\frac{\mbox{\boldmath $B$}^{2}}{8\pi}\mbox{\boldmath $I$}
  \right)
  -\rho\mbox{\boldmath $g$}=0,
\end{eqnarray}
\begin{eqnarray}
  \frac{\partial\mbox{\boldmath $B$}}
       {\partial t}
  = \mbox{\boldmath $\nabla$}
    \times (\mbox{\boldmath $V$}\times\mbox{\boldmath $B$}),
\end{eqnarray}
\begin{eqnarray}
  \frac{\partial}{\partial t}
  \left(
    \rho U
    + \frac{1}{2}\rho\mbox{\boldmath $V$}^{2}
    + \frac{\mbox{\boldmath $B$}^{2}}{8\pi}
  \right)
  +\mbox{\boldmath $\nabla$}\cdot
  \left[
    \left(
      \rho U + p + \frac{1}{2}\rho\mbox{\boldmath $V$}^{2}
    \right)
    \mbox{\boldmath $V$}
    + \frac{c}{4\pi}
      \mbox{\boldmath $E$}\times\mbox{\boldmath $B$}
  \right]
  -\rho\mbox{\boldmath $g$}\cdot\mbox{\boldmath $V$}
  =0,
\end{eqnarray}
and
\begin{eqnarray}
  U=\frac{1}{\gamma-1}
    \frac{p}{\rho},
\end{eqnarray}
\begin{eqnarray}
  \mbox{\boldmath $E$}
  =-\frac{1}{c}
   \mbox{\boldmath $V$}\times\mbox{\boldmath $B$},
\end{eqnarray}
\begin{eqnarray}
  p=\frac{k_{\rm B}}{m}\rho T,
  \label{eq:eos}
\end{eqnarray}
where $U$ is the internal energy per unit mass,
$\mbox{\boldmath $I$}$
the unit tensor,
$k_{\rm B}$ the Boltzmann constant,
$m$ the mean molecular mass,
and $\mbox{\boldmath $g$}=(0,0,-g_{0})$
the
uniform
gravitational acceleration.
Other symbols have their usual meanings
($\rho$ is for density,
$\mbox{\boldmath $V$}$ velocity vector,
$p$ pressure,
$\mbox{\boldmath $B$}$ magnetic field,
$c$ speed of light,
$\mbox{\boldmath $E$}$ electric field,
and $T$ temperature).

To make above equations dimensionless,
we introduce normalizing units of
length $H_{0}$, velocity $C_{{\rm s} 0}$,
time $\tau_{0}\equiv H_{0}/C_{{\rm s} 0}$,
and density $\rho_{0}$,
where $H_{0}=k_{\rm B}T_{0}/(mg_{0})$
is the pressure scale height,
$C_{{\rm s} 0}$ the sound speed,
and $\rho_{0}$ the density at the photosphere,
respectively.
The gas pressure, temperature,
magnetic field strength,
and energy
are normalized by the combinations
of the units above,
i.e., $p_{0}=\rho_{0}C_{{\rm s} 0}^{2}$,
$T_{0}=mC_{{\rm s} 0}^{2}/(\gamma k_{\rm B})$,
$B_{0}=(\rho_{0}C_{{\rm s} 0}^{2})^{1/2}$,
and $E_{0}=\rho_{0}C_{{\rm s} 0}^{2}H_{0}^{3}$,
respectively.
The gravity is given as
$g_{0}=C_{{\rm s} 0}^{2}/(\gamma H_{0})$
by definition.
For comparison of numerical results
with observations,
we use $H_{0}=170\ {\rm km}$,
$C_{{\rm s} 0}=6.8\ {\rm km\ s}^{-1}$,
$\tau_{0}=H_{0}/C_{{\rm s} 0}=25\ {\rm s}$,
and $\rho_{0}=1.4\times 10^{-7}\ {\rm g\ cm}^{-3}$,
which are typical values for the solar photosphere.
Then, $p_{0}=6.3\times 10^{4}\ {\rm dyn\ cm}^{-2}$,
$T_{0}=5600\ {\rm K}$,
$B_{0}=250\ {\rm G}$,
and $E_{0}=3.1\times 10^{26}\ {\rm erg}$.

The initial background stratification
consists of three regions:
an adiabatically stratified convective layer,
a cool isothermal photosphere/chromosphere
(afterward, we simply call it photosphere),
and a hot isothermal corona.
The photosphere and the corona
are smoothly connected by the transition region.
We take $z/H_{0}=0$ to be
the base height of the photosphere,
and the starting height
of the transition region and the corona
are $z_{\rm tr}/H_{0}=10$ and $z_{\rm cor}/H_{0}=20$,
respectively.
The initial temperature
of the photosphere and the corona
are $T_{\rm ph}/T_{0}=1$ and $T_{\rm cor}/T_{0}=150$,
respectively.
The background temperature distribution
in the transition region
is given by
\begin{eqnarray}
  \frac{T_{\rm s}(z)}{T_{0}}
  = \left(
     \frac{T_{\rm cor}}{T_{\rm ph}}
    \right)
    ^{(z-z_{\rm tr})/(z_{\rm cor}-z_{\rm tr})},
\end{eqnarray}
and that of the convection zone being
\begin{eqnarray}
  \frac{T_{\rm s}(z)}{T_{0}}
  = 1 - \frac{z}{T_{0}}
    \left|
     \frac{dT}{dz}
    \right|_{\rm ad},
\end{eqnarray}
where
\begin{eqnarray}
  \left|
   \frac{dT}{dz}
  \right|_{\rm ad}
  = \frac{\gamma-1}{\gamma}\frac{mg_{0}}{k_{\rm B}}
\end{eqnarray}
is the adiabatic temperature gradient
(subscript s is for surrounding distribution).
The initial gas pressure and density profiles
are defined by solving
one-dimensional hydrostatic equation
\begin{eqnarray}
  \frac{d}{dz} p_{\rm s}(z) + \rho_{\rm s}(z) g_{0}=0
\end{eqnarray}
and the equation of state (\ref{eq:eos})
on the basis of the temperature distribution above.

The initial magnetic flux tube
is embedded in the convection zone
at $z_{\rm tube}/H_{0}=-10$.
The longitudinal and azimuthal component
of the flux tube are described as follows:
for a radial distance from the axis
$r=[(y-y_{\rm tube})^{2}+(z-z_{\rm tube})^{2}]^{1/2}$,
\begin{eqnarray}
  B_{x}(r)=B_{\rm tube}
   \exp{\left(-\frac{r^{2}}{R_{\rm tube}^{2}}\right)},
   \label{eq:b1}
\end{eqnarray}
and
\begin{eqnarray}
  B_{\phi}(r)=qrB_{x}(r),
  \label{eq:b2}
\end{eqnarray}
where $(y_{\rm tube},z_{\rm tube})=(0,-10H_{0})$
is the tube center,
$R_{\rm tube}$ the radius,
$q$ the twist parameter,
and $B_{\rm tube}$
the magnetic field strength at the axis.
We take $R_{\rm tube}/H_{0}=2.5$
and $B_{\rm tube}/B_{0}=15$,
i.e., these parameters are
almost the same as those of \citet{mur06}.
For pressure balance
between the flux tube and the surrounding medium,
the gas pressure inside the tube is obtained as
$p_{\rm i}=p_{\rm s}+\delta p_{\rm exc}$,
where
\begin{eqnarray}
  \delta p_{\rm exc}=
  \frac{B_{x}^{2}(r)}{8\pi}
   \left[
     q^{2}\left(
      \frac{R_{\rm tube}^{2}}{2} - r^{2}
      \right)
     -1
   \right].
\end{eqnarray}
The density inside the tube is also defined as
$\rho_{\rm i}=\rho_{\rm s}+\delta\rho_{\rm exc}$,
where
\begin{eqnarray}
  \delta\rho_{\rm exc}
  =\frac{\delta p_{\rm exc}}{p_{\rm s}}
    \rho_{\rm s}
    \exp{\left(-\frac{x^{2}}{\lambda^{2}}\right)},
\end{eqnarray}
and $\lambda/H_{0}=20$.
That is, the flux tube is most buoyant
at the middle of the tube ($x/H_{0}=0$),
and the buoyancy diminishes as $|x|$ increases.

\begin{figure}
  \begin{center}
    \FigureFile(134.4mm,96mm){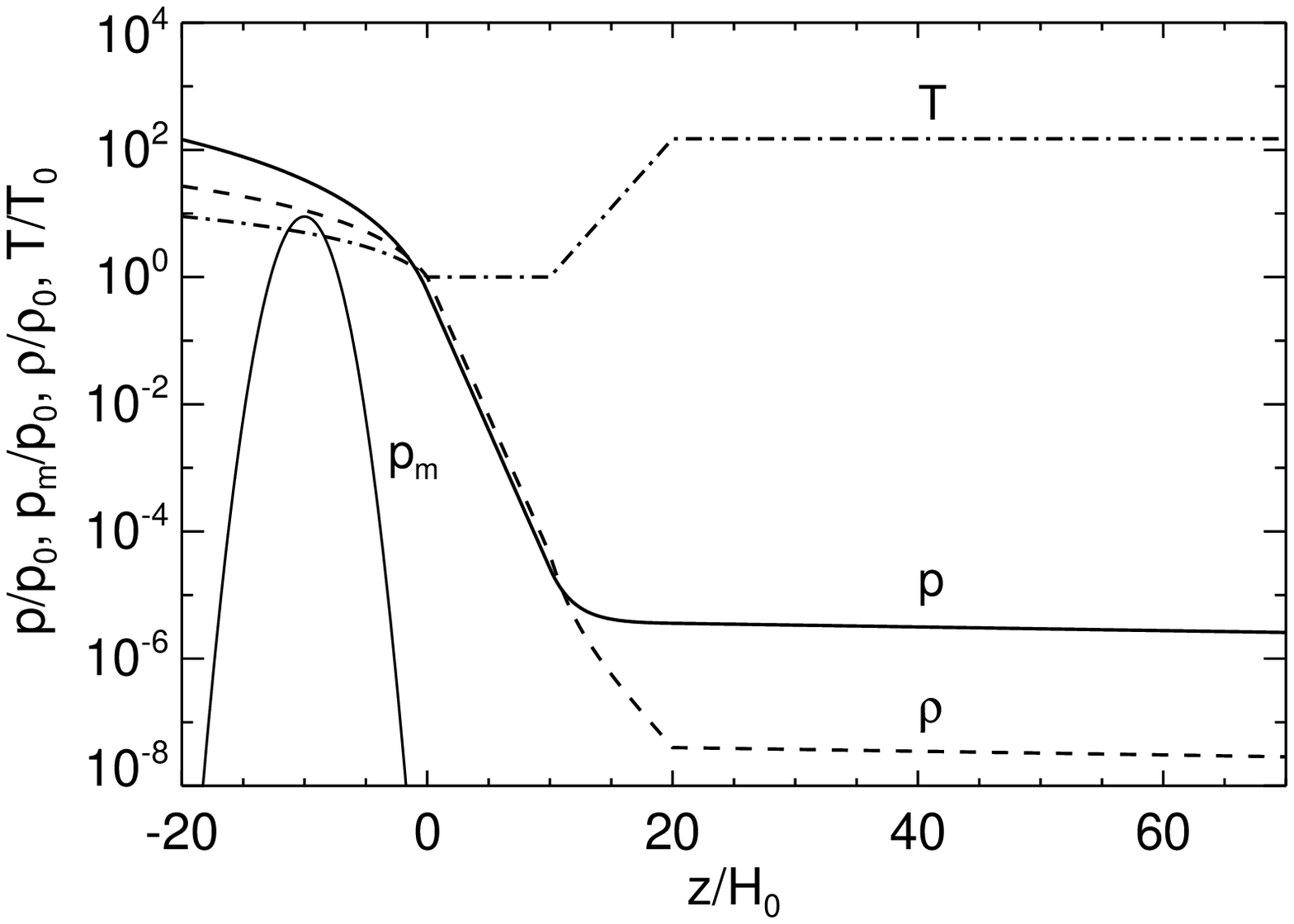}
  \end{center}
\caption{One-dimensional ($z$-)distributions
 of the initial background density (thick solid line),
 pressure (dotted line), and temperature (dashed line).
 The magnetic pressure $p_{\rm m}=B^{2}/(8\pi)$
 of the tube $qH_{0}=0.2$
 along the vertical axis $x/H_{0}=y/H_{0}=0$
 is overplotted with a thin solid line.}
\label{fig:initial}
\end{figure}

Here, we investigate ten parameters of $q$,
which are $qH_{0}=0.5$, 0.4, 0.3, 0.25, 0.2,
0.175, 0.15, 0.125, 0.1, and 0.05.
The plasma beta ($\beta\equiv 8\pi p/B^{2}$)
at the tube center 
is $\beta\sim 3$
at the initial state.
The initial background stratification
(gas pressure, density, and temperature)
and the magnetic pressure
along $x/H_{0}=y/H_{0}=0$
of the case $qH_{0}=0.2$
are indicated in Figure \ref{fig:initial}.

The simulation domain is taken as
$(-120,\ -120,\ -20)\le(x/H_{0},\ y/H_{0},\ z/H_{0})
\le(120,\ 120,\ 150)$,
resolved by $256\times 256\times 256$ grids.
The grid spacings
for $x$, $y$, and $z$ directions
are $\Delta x/H_{0}=\Delta y/H_{0}=0.5$
for $(-40,\ -40)\le(x/H_{0},\ y/H_{0})\le(40,\ 40)$,
and $\Delta z/H_{0}=0.2$
for $-20\le z/H_{0}\le 20$,
respectively.
Outside this range,
the mesh sizes gradually increase.
We assume periodic boundaries
for horizontal directions
and symmetric for vertical.
A wave-damping region is attached
near the top boundary.
We use the modified Lax-Wendroff scheme version of the CANS
(Coordinated Astronomical Numerical Software) code
(see \cite{tor10a}).

\section{Results\label{sec:results}}

\subsection{Overview of the Results\label{sec:overview}}

\begin{figure}
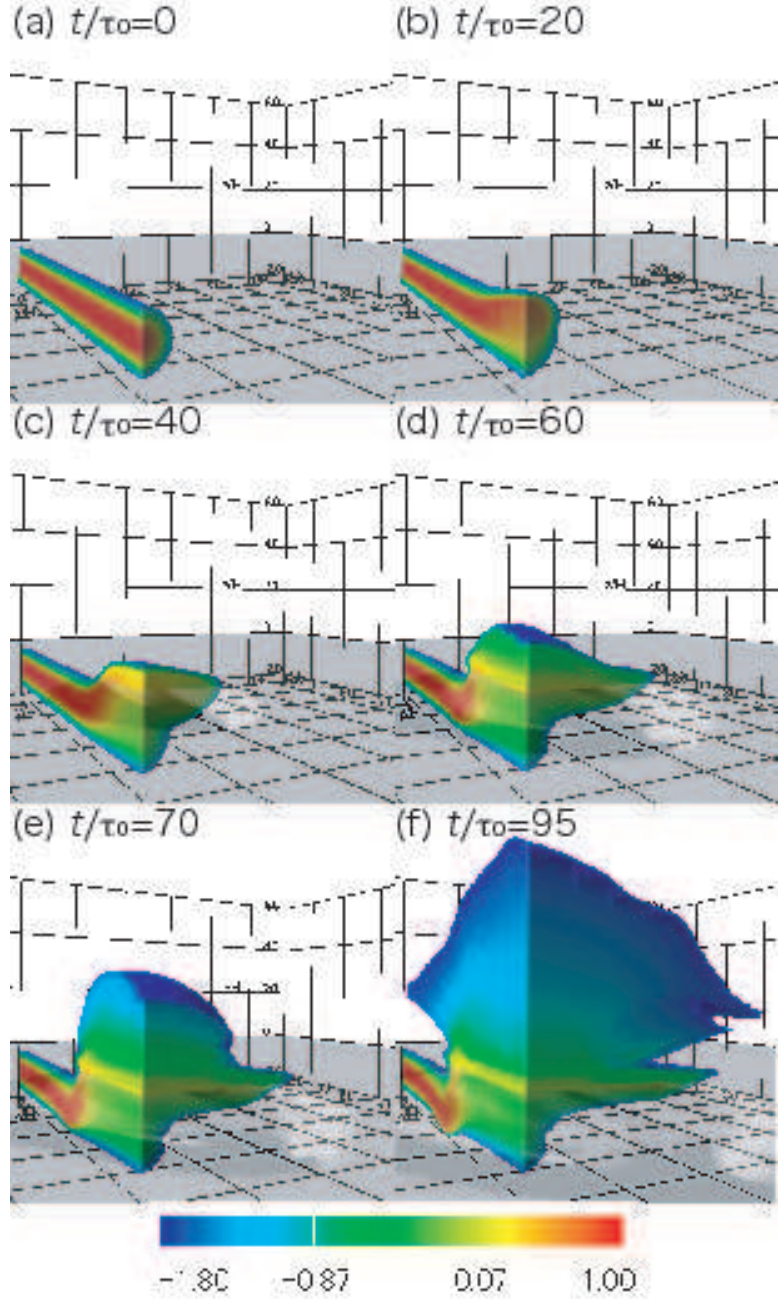

\begin{center}

\FigureFile(102mm,17.35mm){figure2.eps2}

\end{center}

\caption{Time evolution of the flux tube with the twist $qH_{0}=0.2$.
Logarithmic field strength $\log{(|B|/B_{0})}$
and photospheric magnetogram $B_{z}/B_{0}$
are plotted.
In each panel,
the region $x/H_{0}\le 0$ and $y/H_{0}\ge 0$ is shown.
This figure is also available as an avi animation in the electronic
edition.}
\label{fig:general}
\end{figure}

Figure \ref{fig:general} shows the time evolution
of the flux tube with the twist $qH_{0}=0.2$.
In this Figure,
we plot the logarithmic field strength
$\log{(|B|/B_{0})}$
and the photospheric magnetogram $B_{z}/B_{0}$.
In each panel,
the region $x/H_{0}\le 0$ and $y/H_{0}\ge 0$ is shown.
Initially, the flux tube is embedded
at $z_{\rm tube}/H_{0}=-10$,
and is slightly buoyant
around the tube center $-20<x/H_{0}<20$.
The tube rises
through
the convection zone
by magnetic buoyancy,
and reaches the surface at $t/\tau_{0}=20$
(Figure \ref{fig:general}(b)),
while the outskirt of the rising portion
($x/H_{0}\sim -20$) begins to sink,
because the fluid is drained
along the field lines
from the apex of the rising tube.
Due to the isothermal
(i.e., strongly-subadiabatic) photosphere,
the tube is decelerated
and expands laterally near the surface
to make a ``photospheric tongue''
(Figure \ref{fig:general}(c)).
That is, the convectively stable photosphere
inhibits an upward motion of the fluid,
and thus,
the magnetic field cannot penetrate the photosphere
only to escape in the horizontal direction.
(It should be noted that the term ``tongue'' here
is different from that used
in \citet{li07} and \citet{arc10}.
The horizontal extension
of the magnetic field
is also observed
in a recent radiative MHD calculation
by \citet{che10}.
The lateral expansion speed is
$|V_{y}|\sim 0.4C_{{\rm s}0}=2.7\ {\rm km\ s}^{-1}$,
i.e., a fraction of the photospheric sound speed.
At this moment,
the photospheric magnetogram
shows a north-south ($y$-directional)
magnetic distribution,
and the total field strength is
$|B|\sim 2B_{0}=500\ {\rm G}$
and plasma beta is $\beta\equiv p/p_{\rm mag}\sim 2$
around the photosphere.

As the magnetic pressure gradient enhances,
the second-step emergence takes place.
In Figure \ref{fig:general}(d),
multiple expansions are observed:
the rise velocity is
about $(0.3-0.5)C_{{\rm s}0}=2.0-3.4\ {\rm km\ s}^{-1}$.
This multi-dome structure is more noticeable
in a weaker twist case.
Figure \ref{fig:ro-mag}(a) shows
the magnetic field structure
of the case $qH_{0}=0.15$ at $t/\tau_{0}=80$.
The corresponding density structure
and velocity vectors
in the $y/H_{0}=0$ plane,
and the field lines
are indicated
in Figure \ref{fig:ro-mag}(b).
From these figures,
one can see four magnetic domes are built
and the fluid is accumulated
between the domes.
In this region,
the field lines are generally
directed in the $y$-direction.
Therefore, this situation can be explained
as a consequence
of the interchange mode
of the magnetic buoyancy instability,
i.e., the wavenumber vector is perpendicular
to the field lines.

As time goes on,
the flux tube expands
both vertically and horizontally,
while the photospheric tongue
also continues to expand laterally
(Figure \ref{fig:general}(e)).
Finally, the flux tube makes a single dome
of the height $z/H_{0}\sim 60$
on the pancake-like structure
at the surface
(Figure \ref{fig:general}(f)).
At the same time,
the sunk part approaches
the bottom of the simulation domain
($z/H_{0}\sim -20$).

Overall evolution described above
is similar to the observation by \citet{ots10}.
They found the lateral expansion
with the speed of $2.9\ {\rm km\ s}^{-1}$
at the surface
before further evolution occurred,
which is consistent with our results
of $|V_{y}|=2.7\ {\rm km\ s}^{-1}$.
The gradual rise speed of the secondary emergence
was observed to be $2.1\ {\rm km\ s}^{-1}$,
which is also consistent with our results
of $2.0-3.4\ {\rm km\ s}^{-1}$.

\subsection{Parameter Study on the Twist Strength
\label{sec:param}}

\begin{figure}
  \begin{center}
    \FigureFile(134.4mm,96mm){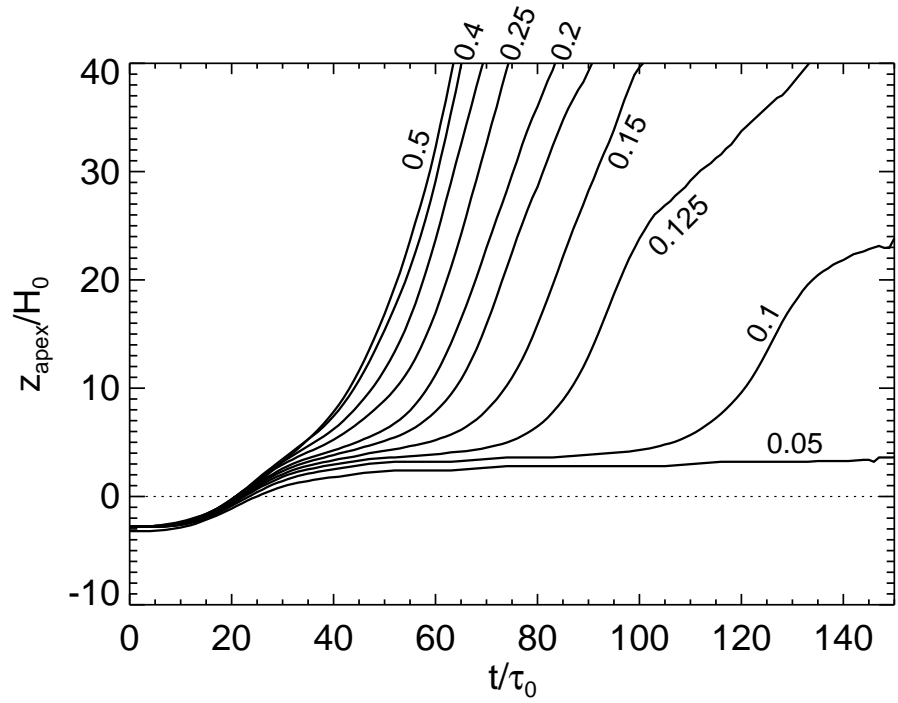}
  \end{center}
\caption{Height-time relations for various twist cases.
From left to right, each line shows the evolution of
$qH_{0}=0.5$, 0.4, 0.3, 0.25, 0.2, 0.175, 0.15, 0.125,
0.1, and 0.05.
It takes more time to rise into the corona
as the twist strength decreases.}
\label{fig:param}
\end{figure}

Figure \ref{fig:param} shows
the height-time relation of the top of the tube
for various twist cases.
Here, in this figure,
we plot the height at the highest portion
of the emerging flux tube
($z_{\rm apex}/H_{0}$).
The evolutions are found
to depend on the initial twist,
and each line distributes in a continuous fashion.
It can be seen from this figure
that almost all the tubes
($0.5\ge qH_{0}\ge 0.1$)
show the two-step emergence to the corona.
Although the rise times
within the convection zone
are similar to each other,
tubes with weaker twists spend more time
in the surface
waiting for the second-step emergence
to be triggered.
The tube with $qH_{0}=0.1$
shows only a slight emergence
in the atmosphere
($z/H_{0}\lesssim 20$).
As for the weakest twist case with $qH_{0}=0.05$,
further evolution never takes place
within the elapse calculated
(failed emergence).
These results are consistent with those of
\citet{mag01}, \citet{mur06}, and \citet{tor10b}.

\begin{figure}
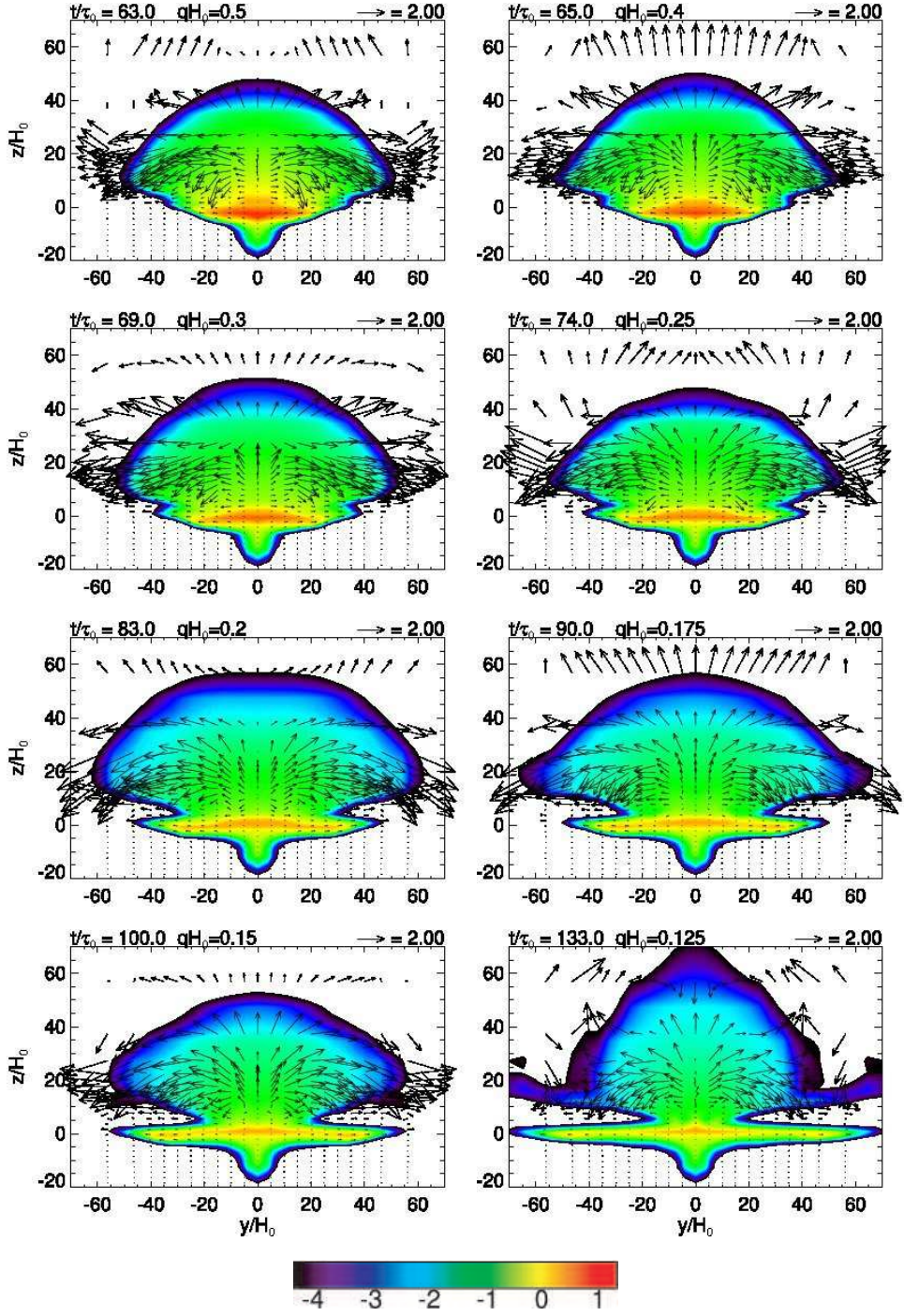

\begin{center}

\FigureFile(65.25mm,40.5mm){figure5a.eps2}
\FigureFile(60.75mm,40.5mm){figure5b.eps2}\\
\FigureFile(65.25mm,40.5mm){figure5c.eps2}
\FigureFile(60.75mm,40.5mm){figure5d.eps2}\\
\FigureFile(65.25mm,40.5mm){figure5e.eps2}
\FigureFile(60.75mm,40.5mm){figure5f.eps2}\\
\FigureFile(65.25mm,43.5mm){figure5g.eps2}
\FigureFile(60.75mm,43.5mm){figure5h.eps2}\\
\FigureFile(46.05mm,6.75mm){figure5i.eps2}

\end{center}

\caption{
Cross-sections at $x/H_{0}=0$ plane
of eight out of ten flux tubes
that reach $z/H_{0}=40$.
Each figure shows the logarithmic field strength
$\log{(|B|/B_{0})}$
when the tube reaches that height.
}
\label{fig:cross}
\end{figure}

Cross-sections at $x/H_{0}=0$ plane
of eight out of ten flux tubes
are shown in Figure \ref{fig:cross}.
These tubes are those who reach $z/H_{0}=40$,
and each figure shows the arrival at that height.
As the initial twist $qH_{0}$ becomes smaller,
the lateral expansion at the surface
(tongue-like structure
around $z/H_{0}\sim 0$)
is increasingly remarkable,
because the more intense initial twist
of the magnetic flux tube
yields the stronger azimuthal
magnetic tension force,
and thus keeps the tube coherent.
Therefore,
the coronal field intensity
also reduces with decreasing $qH_{0}$,
and the coronal structure with a weaker twist
is more fragmented
compared to the stronger twist cases.

\subsection{Comparison of the Magnetic Field Structure
at the Surface\label{sec:compare}}

\begin{figure}
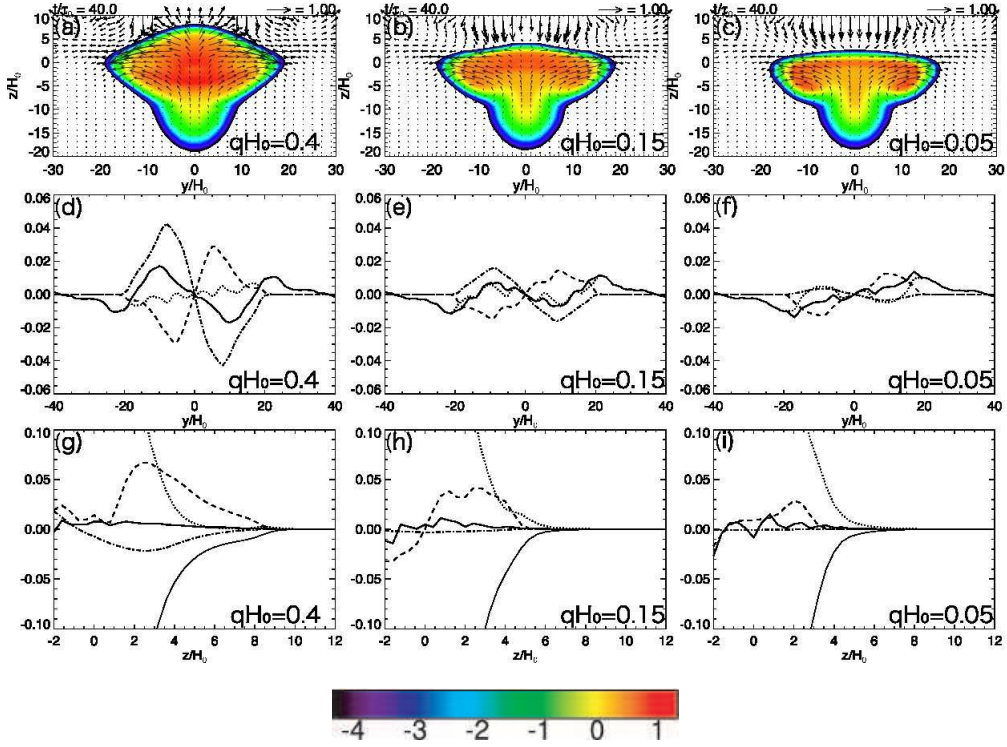

\begin{center}

\FigureFile(132.64mm,87.84mm){figure6a.eps2}\\
\FigureFile(46.05mm,6.75mm){figure6b.eps2}

\end{center}

\caption{
({\it Top}) Cross-sectional configuration
of the flux tubes at $x/H_{0}=0$
with velocity vectors,
({\it middle}) horizontal components of each force
along the horizontal axis $x/H_{0}=z/H_{0}=0$,
and ({\it bottom}) vertical components of forces
along the vertical axis $x/H_{0}=y/H_{0}=0$
for cases with $qH_{0}=0.4$ (rapid emergence),
0.15 (slow), and 0.05 (failed)
at the time $t/\tau_{0}=40$.
Plotted lines are the total force (thick solid line),
gas pressure gradient (dotted),
magnetic pressure gradient (dashed),
magnetic tension (dash-dotted),
and gravity (thin solid),
respectively.
}
\label{fig:compare}
\end{figure}

In this subsection,
we compare the magnetic field structures
of the different emergence cases
to study the horizontal expansion
and the mechanism of the second-step evolution
at the solar surface.
Figure \ref{fig:compare}
shows ({\it top}) the cross-sectional configuration
of the flux tubes at $x/H_{0}=0$
with velocity vectors,
({\it middle}) the horizontal components of forces
along the horizontal axis $x/H_{0}=z/H_{0}=0$,
and ({\it bottom}) the vertical components of forces
along the vertical axis $x/H_{0}=y/H_{0}=0$
for cases with $qH_{0}=0.4$ (rapid emergence),
0.15 (slow emergence), and 0.05 (failed emergence)
at the time $t/\tau_{0}=40$.

As can be seen from the top figures,
the second-step evolution has already begun
at this time for the case with $qH_{0}=0.4$,
while the vertical expansion cannot be seen
in $qH_{0}=0.05$ case.
It should be noted that
the downflow in the uppermost areas
in Figures \ref{fig:compare}(b) and (c)
is a reflected wave from the top boundary.
Horizontally,
for a strongest twist case
(Figure \ref{fig:compare}(d)),
an inward magnetic tension force is dominant,
and thus the inward total force keeps
the tube from a lateral fragmentation.
As the twist decreases,
the magnetic tension reduces
so that the total force is outward
in a wider range
for $qH_{0}=0.05$
(Figure \ref{fig:compare}(f)),
resulting in the tube's fragmentation
and the further expansion never to occur.
As for vertical forces,
magnetic pressure gradient is principal
at front of the tube
(Figure \ref{fig:compare}(g): $3<z/H_{0}<9$).
However,
the total force is about zero
within this area.
That is, the second-step expansion
is caused by the tube's magnetic pressure,
while the tube is almost in a hydrostatic equilibrium
with surrounding materials.
For a weak twist case
(Figure \ref{fig:compare}(i)),
the magnetic pressure gradient
is much less effective.
Therefore, further rise cannot occur.

\subsection{Undulating Configuration
of the Photospheric Field Lines\label{sec:bvct}}

Figure \ref{fig:bvct} shows
the surface magnetogram $B_{z}/B_{0}$
and the field lines above the surface
for the medium twist case ($qH_{0}=0.15$)
at the time $t/\tau_{0}=125$,
namely, in the later phase.
We confirm that some undulating field lines
connect magnetic patches at the surface,
and that, as time goes on,
they gradually rise into the corona
by forming longer fields.
At this time,
both near-surface undulating fields
and coronal fields
are directed almost parallel
to the axis of the original flux tube,
which is in contrast to the perpendicular fields
observed in the earlier phase of the emergence
(see Figure \ref{fig:ro-mag}(b)).
This result is reminiscent of
the ``sea-serpent'' field lines
and the resistive emergence
by \citet{par04}.
Undulating photospheric fields
of a weakly twisted tube
are also found by \citet{arc10}.
We will discuss this again
in Section \ref{sec:discussion}.

\subsection{Magnetic Energy in the Atmosphere
and the Initial Tube's Twist
\label{sec:energy}}

\begin{figure}
  \begin{center}
    \FigureFile(134.4mm,96mm){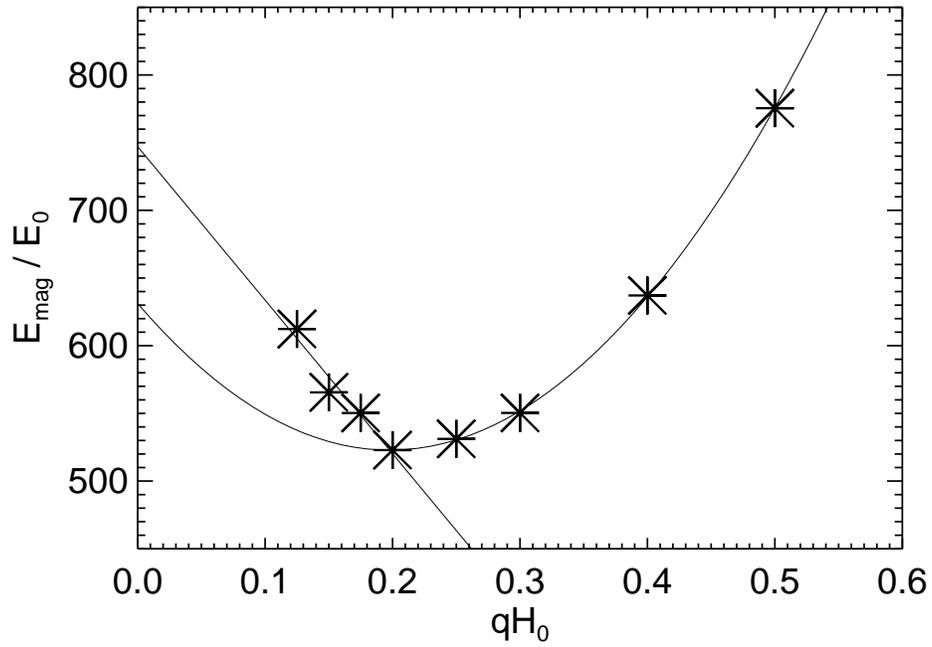}
  \end{center}
\caption{
Total magnetic energy above the surface
$E_{\rm mag}=\int_{z>0} B^{2}/(8\pi)\, dV$
versus the initial twist $q$,
measured at the time
when each tube reaches $z/H_{0}=40$.
Quadratic and linear lines are overplotted
with solid lines.
}
\label{fig:energy}
\end{figure}

Figure \ref{fig:energy}
shows the initial twist $qH_{0}$
and the total magnetic energy
measured above the solar surface $E_{\rm mag}/E_{0}$
when each tube arrives at $z/H_{0}=40$
(see Figure \ref{fig:cross}).
The magnetic energy above the surface
is defined as:
\begin{eqnarray}
  E_{\rm mag}
  = \int_{z>0} \frac{B^{2}}{8\pi}\ dV.
\end{eqnarray}

When the initial twist is large ($qH_{0}\ge 0.2$),
the magnetic energy $E_{\rm mag}$
is found to obey a $q^{2}$ law.
From the field configuration
(\ref{eq:b1}) and (\ref{eq:b2}),
the initial magnetic energy per unit volume
can be calculated as
\begin{eqnarray}
  \frac{B^{2}}{8\pi}
  = \left(q^{2}r^{2}+1\right)
    \frac{B_{\rm tube}^{2}}{8\pi}
    \exp{\left(-\frac{2r^{2}}{R_{\rm tube}^{2}}\right)},
\end{eqnarray}
that is,
the initial tube's magnetic energy
depends on $q^{2}$.
When the twist is strong,
the azimuthal magnetic tension force
is more effective
and the lateral expansion around the photosphere
is less efficient
that the initial magnetic energy
is directly transported
into the atmosphere above the surface.
As a result of this regime,
the observed magnetic energy
within the atmosphere
relies on a $q^{2}$ function.

Contrary to this,
when the initial twist is weak
($qH_{0}\le 0.2$),
the consequent magnetic energy
deviates from the $q^{2}$ line
and is negatively-correlated
with the initial twist.
It is speculated that,
for these tubes with weaker twits,
it takes a longer time
to reach $z/H_{0}=40$
from the surface.
It is because
the photospheric field needs more time
to satisfy the condition for the second-step emergence
(i.e. \cite{ach79}),
for the weak azimuthal field
causes the tube to expand horizontally
around the surface.
Since the magnetic energy is
continuously transported from below
and the horizontal expanding velocity
is almost the same for $qH_{0}\le 0.2$ cases
($V_{y}/C_{{\rm s} 0}\sim 0.1$),
the magnetic energy of the photospheric tongue
is expected to depend on the time lag
between the tube's arrival at the photosphere
and at $z/H_{0}=40$.

It can be concluded that
the magnetic energy in the atmosphere
correlates critically with the initial twist.
In the strong twist regime,
the energy-twist relation
follows the initial $q^{2}$ rule,
because the tightly twisted tube
does not exhibit a significant expansion
near the photosphere.
As for the weak twist cases,
they depend on the time lag
between reaching the photosphere and the corona.
It is because
the weaker twist tube
takes more time
to rise further,
and, therefore,
more magnetic energy is stored
in the photospheric tongue.

\section{Summary\label{sec:summary}}

In this paper,
we carried out three-dimensional MHD simulations
to investigate the effect of the initial twist
on the flux tube evolution.
Here, we summarize the results:
\begin{itemize}
\item Initially, the flux tube rises
through the convection zone
due to its magnetic buoyancy.
Reaching the surface,
the tube expands laterally
to make a ``photospheric tongue.''
The secondary emergence occurs
after sufficient flux accumulates
within the photosphere.
Due to the interchange mode instability,
the tube builds multiple domes above the surface,
between which the fluids piles up.
Finally, the flux tube arrives at
$z/H_{0}\sim 60$
as a single dome.
The overall emergence is consistent
with the recent observations
(e.g. \cite{ots10}).

\item We run ten twist cases
to investigate the effect of the initial twist.
Nine out of ten reach the coronal height
($z/H_{0}\ge20$)
showing two-step emergence,
while the weakest twist case fails
to rise further above the surface
($qH_{0}=0.05$).
In the two-step emergence regime,
the rise time becomes shorter
with increasing initial twist,
which is consistent with the previous
calculations by \citet{mur06}.
The photospheric tongue is
more noticeable in weaker twist case.

\item We study the force components
at the solar surface
for different twist cases
at the time $t/\tau_{0}=40$.
The stronger the initial twist is,
the larger the inward magnetic tension is,
resulting the tube keeps its coherency.
In the weak twist case,
the magnetic tension is much less effective,
causing the tube distorted.
At the same time,
the strong twist tube rises
further into the atmosphere
mainly by the magnetic pressure gradient.

\item We found that the photospheric fields
of the middle twist case ($qH_{0}=0.15$)
undulate in the later phase of the emergence.
The field lines
gradually rise into the corona
as longer loops.
The photospheric and coronal fields are
almost parallel to the axis
of the initial flux tube.
These features remind us
of the resistive emergence model
by \citet{par04}.

\item We measure the magnetic energy
$E_{\rm mag}$ above the surface.
The energy plot follows the initial $q^{2}$ law
when the twist is strong ($qH_{0}\ge 0.2$),
while, for weaker twist cases ($qH_{0}\le 0.2$),
the energy depends on
the time difference
between reaching the surface and the corona.
That is,
weakly twisted tube takes more time
for magnetic flux to accumulate near the surface
and the secondary instability to be triggered.
\end{itemize}

\section{Discussion\label{sec:discussion}}

In Section \ref{sec:results},
we showed the time-evolution
of the twisted flux tube.
When the second-step emergence starts,
multiple domes are observed
above the surface
and fluid is trapped
between the expanding magnetic structures
(see Figure \ref{fig:ro-mag}).
At this time,
field lines are directed
perpendicular to the alignment
of the magnetic domes.
For middle twist tube,
we also found undulating fields
near the surface,
emerging into the corona
(see Figure \ref{fig:bvct}).
In this section,
we discuss these features
in connection with future observations.

\subsection{Twist Intensity and the Interchange Instability
\label{sec:interchange}}

In Section \ref{sec:overview},
we saw that,
as the twist decreases,
the interchange mode instability
becomes more noticeable.
However, this is contrary to the expectation
that the azimuthal field
should be less pronounced
in a weaker twist case.
It may be because,
in a weak twist case,
the tube extends laterally near the photosphere
and thus the twist increases.
As the tube develops
the interchange instability,
field lines perpendicular
to the alignment of the magnetic domes
become more pronounced
(see Sections \ref{sec:param} and \ref{sec:compare}).

\subsection{Twist of the Actual Flux Tube in the Sun
\label{sec:twist}}

Multiple magnetic structures
and the density accumulation between them
are also found
in previous observations and calculations.
\citet{par04} found
that photospheric fields are undulating
at its earlier phase of the flux emergence event,
and proposed a resistive emergence model
that undulating multiple loops reconnect
with each other
to make larger coronal fields.
\citet{iso07} carried out
two-dimensional MHD simulation
to study the evolution
of the serpentine magnetic loops
(resistive emergence model),
finding that density accumulates
in between the magnetic loops;
their elongated vertical plasma structures
are similar to our results
(for three-dimensional study, see \citet{arc09}).
However, in our model,
field lines are directed
almost perpendicular
to the alignment of the domes,
which is against the observations
of the undular field lines
(e.g. \cite{par04}).
The direction of the field lines
are the consequence of the initial tube's twist.
Therefore, the difference between
the present calculations and the observations
indicates that
the actual twist of the flux tube
beneath the surface
may be much weaker than those assumed in our models
(e.g. $qH_{0}=0.15$ for Figure \ref{fig:ro-mag}).

On the other hand,
flux tube with insufficient twist
was found to fail
to rise through the convection zone
\citep{mor96,emo98,tor10b}.
It is because
the weak azimuthal field
of the flux tube
cannot hold its coherency
during its ascent
within the solar interior.
Therefore,
one of the important problems
to be solved
is the emergence of the flux tube
with much weaker twist
($qH_{0}\le 0.1$).

We also found that
the medium twist tube ($qH_{0}=0.15$) reveals
the undulating fields at the surface,
which gradually rise into the upper atmosphere
as longer coronal loops
(Section \ref{fig:bvct}).
These fields are directed
parallel to the main axis
of the initial flux tube.
This picture seems well accorded
with the resistive emergence model.
However, it is in the later phase
that this undulatory evolution is observed,
and, in the earlier phase,
the field lines are perpendicular
to the original tube's axis
(see Figure \ref{fig:ro-mag}).
Therefore,
the reproduction of the undulating fields
parallel to the original axis
\citep{par04}
is not achieved.

\subsection{Formation
of Undulating Photospheric Fields
\label{sec:undulation}}

Recently, \citet{che10} have conducted
a radiative MHD simulation
of the formation of an active region.
They showed that
the rising flux tube flattens
to make a pancake-like structure
near the surface,
and that the convective flows
create serpentine field lines.

In the present study,
which does not include the convection,
we also found
the sideway expansion of the rising field
at the photosphere (tongue)
and the undulation of the photospheric fields
in the later phase of the emergence
of the weaker twist tube.
These features are also confirmed
by \citet{arc10};
their simulations do not
take account of the convection effects,
either.
Therefore, we can see
that other mechanisms,
apart from the convection,
could also explain the formation
of serpentine fields at the photosphere.

\subsection{For Future Observations
\label{sec:future}}

In this paper,
we found some aspects
of the flux emergence event.
One is the photospheric tongue,
i.e., the magnetic structure
extending horizontally
around the surface
just before further evolution takes place.
Temporally- and spacially-resolved
spectroscopic observations
of the earlier phase of the flux emergence
are required
to study this magnetic extension
at the photosphere.

At the same time,
we found that the initial twist
of $qH_{0}\sim 0.1$
at $-1700\ {\rm km}$
is too strong
to match the observations
(as mentioned above in Section \ref{sec:twist}).
Local and global helioseismology
are needed to reveal
the flux emergence
(especially on the twist evolution)
within the convection zone.
The key issue
is how weakly twisted flux tubes
manage to rise through the solar interior.

\bigskip

Numerical computations were
carried out on NEC SX-9
at the Center for Computational Astrophysics, CfCA,
of the National Astronomical Observatory of Japan,
and on M System (Fujitsu FX1)
of JAXA Supercomputer System.
The page charge of this paper
is partly supported by CfCA.
S. T. and T. Y. thank Dr. Y. Fan of the High Altitude Observatory,
the National Center for Atmospheric Research.
We thank the referee for helpful suggestions
for improvements of this paper.


\clearpage

\clearpage

\clearpage
\begin{figure}
\begin{center}

\FigureFile(101.67mm,165.33mm){figure3.eps2}

\end{center}

\caption{
The flux tube with $qH_{0}=0.15$ at the time $t/\tau_{0}=80$.
(a) Magnetic field structure $\log{(|B|/B_{0})}$
and photospheric magnetogram $B_{z}/B_{0}$.
Plotted colors are the same as those of Figure \ref{fig:general}.
(b) The corresponding logarithmic density profile
$\log{(\rho/\rho_{0})}$ at $y/H_{0}=0$ plane
($-40\le x/H_{0}\le -5$ and $5\le z/H_{0}\le 17.5$)
with velocity vectors (white arrows),
magnetogram at $z/H_{0}=5$,
and field lines (blue lines) are shown.
}
\label{fig:ro-mag}
\end{figure}

\clearpage

\clearpage

\clearpage

\clearpage
\begin{figure}
  \begin{center}
    \FigureFile(100mm,100mm){figure7.eps2}
  \end{center}
\caption{
Surface magnetogram $B_{z}/B_{0}$
(color contour)
and the field lines (blue lines)
for the middle twist case ($qH_{0}=0.15$)
at the time $t/\tau_{0}=125$ are shown.
Note that some filed lines undulate near the surface
and gradually rise into the corona
as longer loops.
}
\label{fig:bvct}
\end{figure}

\clearpage

\end{document}